\documentclass[8.5pt,twoside,twocolumn]{article}
\usepackage[super,sort&compress,comma]{natbib} 
\usepackage[version=3]{mhchem}
\usepackage{balance}
\usepackage{times,mathptm}
\usepackage{sectsty}
\usepackage{graphicx} 
\usepackage{lastpage}
\usepackage[format=plain,justification=raggedright,singlelinecheck=false,font=small,labelfont=bf,labelsep=space]{caption} 
\usepackage{fancyhdr}
\begin{document}

\thispagestyle{plain}
\fancypagestyle{plain}{
\renewcommand{\headrulewidth}{1pt}}
\renewcommand{\thefootnote}{\fnsymbol{footnote}}
\renewcommand\footnoterule{\vspace*{1pt}%
\hrule width 3.4in height 0.4pt \vspace*{5pt}} 
\setcounter{secnumdepth}{5}

\makeatletter 
\renewcommand\@biblabel[1]{#1}            
\renewcommand\@makefntext[1]%
{\noindent\makebox[0pt][r]{\@thefnmark\,}#1}
\makeatother 
\renewcommand{\figurename}{\small{Fig.}~}
\sectionfont{\large}
\subsectionfont{\normalsize} 

\fancyfoot{}
\fancyfoot[RO]{\footnotesize{\sffamily{\textbar  \hspace{2pt}\thepage}}}
\fancyfoot[LE]{\footnotesize{\sffamily{\thepage~\textbar}}}
\fancyhead{}
\renewcommand{\headrulewidth}{1pt} 
\renewcommand{\footrulewidth}{1pt}
\setlength{\arrayrulewidth}{1pt}
\setlength{\columnsep}{6.5mm}
\setlength\bibsep{1pt}

\twocolumn[
  \begin{@twocolumnfalse}
\noindent\LARGE{\textbf{Absolute density measurement of SD radicals in a supersonic jet at the quantum-noise-limit}}
\vspace{0.6cm}

\noindent\large{\textbf{Arin Mizouri,\textit{$^{a}$} Lianzhong Deng,\textit{$^{b,c}$} Jack S. Eardley,\textit{$^{b}$} N. Hendrik Nahler,\textit{$^{d}$} Eckart Wrede,\textit{$^{b\ddag}$} and
David Carty$^{\ast}$\textit{$^{a,b\ddag}$}}}\vspace{0.5cm}

\noindent\textit{\small{\textbf{Received Xth XXXXXXXXXX 20XX, Accepted Xth XXXXXXXXX 20XX\newline
First published on the web Xth XXXXXXXXXX 200X}}}

\noindent \textbf{\small{DOI: 10.1039/b000000x}}
 \end{@twocolumnfalse} \vspace{0.6cm}

  ]

\noindent\textbf{The absolute density of SD radicals in a supersonic jet has been measured down to $\mathbf{(1.1\pm0.1)\times10^5}$~cm$\mathbf{^{-3}}$ in a modestly specified apparatus that uses a cross-correlated combination of cavity ring-down and laser-induced fluorescence detection. Such a density corresponds to $\mathbf{215\pm21}$ molecules in the probe volume at any given time. The minimum detectable absorption coefficient was quantum-noise-limited and measured to be $\mathbf{(7.9\pm0.6)\times10^{-11}}$~cm$\mathbf{^{-1}}$, in 200~s of acquisition time, corresponding to a noise-equivalent absorption sensitivity for the apparatus of $\mathbf{(1.6\pm0.1)\times10^{-9}}$~cm$\mathbf{^{-1}}$~Hz$\mathbf{^{-1/2}}$.}

\section*{}
\vspace{-1cm}


\footnotetext{\textit{$^{a}$~Department of Physics, Durham University, South Road, Durham, DH1 3LE, United Kingdom. E-mail: david.carty@durham.ac.uk}}
\footnotetext{\textit{$^{b}$~Department of Chemistry, Durham University, South Road, Durham DH1 3LE, United Kingdom. }}
\footnotetext{\textit{$^{c}$~Present address:  State Key Lab of Precision Spectroscopy, East China Normal University, No.3663 North Zhongshan Road, Shanghai  200062, P.R.China. }}
\footnotetext{\textit{$^{d}$~School of Engineering and Physical Sciences, Heriot-Watt University, Edinburgh EH14 4AS, United Kingdom. }}


\footnotetext{\ddag~These authors contributed equally to this work.}

The knowledge of absolute densities of dilute molecular samples is important in a diverse range of disciplines including atmospheric chemistry and environmental pollutants monitoring,\cite{Stone2012,Fried2008} combustion and automotive engineering,\cite{Rahinov2006,Spearrin2013} astrochemistry,\cite{Blitz2012} plasma physics,\cite{Ionin2007} cold and ultracold molecules,\cite{Wang2010} manufacturing,\cite{Booth2000} radiocarbon dating,\cite{Galli2011} detection of chemical weapons agents and toxic industrial chemicals and explosives\cite{Patel2008} and medical diagnostics and monitoring.\cite{Schmidt2013} In the field of chemical reaction dynamics, the lack of a straightforward and sensitive method for measuring molecular beam densities is currently a significant barrier to determining precise absolute cross sections in scattering experiments.\cite{Kirste2012}

Laser-induced fluorescence (LIF) is a well established technique for sensitively detecting very low densities of suitable species. With careful attenuation of stray light and photon-counting over very long acquisition times, single figure numbers of molecules can be detected in the probe volume. However, it is well known that calibrating a LIF setup to give absolute densities can be very difficult and imprecise. Kirste (and co-workers) recently measured, over nearly 4 hours, densities of OH radicals as low as 200~molecules~cm$^{-3}$ (with an error around 30\%) in their 0.03~cm$^3$ probe volume, obtaining absolute densities by painstakingly calibrating their detection system and the fluorescence process (fluorescence quantum yield, solid angle observed by the detector, detector quantum efficiency and transmission efficiency of optical components in front of the detector).\cite{Kirste2012}

Absorption techniques, such as cavity ring-down spectroscopy (CRDS), are also well-established and widely used for direct quantitative measurements of molecules in gas, liquid and solid-phase samples. In particular absolute absorption coefficients $\alpha=\sigma\rho$ are measured and, therefore, absolute densities $\rho$ can be determined, if the absorption cross-section $\sigma$ is known. The minimum detectable absorption coefficient, $\alpha_\text{min}$, depends on what region of the electromagnetic spectrum is being used; of particular importance being the reflectivity of the cavity mirrors, the bandwidth of lasers and whether the lasers are CW or pulsed. In general, $\alpha_\text{min}$ increases as one goes from the IR, through the visible region, into the UV. As an example of the state-of-the-art in the IR, Foltynowicz \textit{et al}. used the CRDS variant NICE-OHMS  (CW, 1531~nm) to detect \ce{C2H2} (10~ppm in 0.2~mbar of \ce{N2}) achieving an $\alpha_\text{min}=1.3\times10^{-12}$~cm$^{-1}$ in 400~s, \emph{i.e.} a noise-equivalent absorption sensitivity (NEA) of $1.8\times10^{-11}$~cm$^{-1}$~Hz$^{-1/2}$.\cite{Foltynowicz11} In the visible (pulsed, 100~Hz, 532~nm), Osthoff \textit{et al.}\ detected \ce{NO2} achieving an $\alpha_\text{min}=1.7\times10^{-10}$~cm$^{-1}$ in 1~s, a NEA of $2.4\times10^{-10}$~cm$^{-1}$~Hz$^{-1/2}$.\cite{Osthoff06} In the UV (pulsed, 10~Hz, 254~nm), Jongma \textit{et al.} detected Hg with an $\alpha_\text{min}=8.3\times10^{-7}$~cm$^{-1}$ in 3~s, a NEA of $2.0\times10^{-6}$~cm$^{-1}$~Hz$^{-1/2}$.\cite{Jongma95}

Recently, Sanders~\textit{et~al}. described a method called cavity-enhanced laser-induced fluorescence (CELIF) that combines the absolute absorption capabilities of CRDS and the sensitivity of LIF to measure absolute absorption coefficients.~\cite{Sanders2013} They showed that the technique was particularly effective for molecules like 1,4-bis(phenylethynyl)benzene (BPEB), which has a short fluorescence lifetime (500~ps), in a supersonic jet in the UV (pulsed, 10~Hz, 320~nm). An $\alpha_\text{min}<1.5\times10^{-9}$~cm$^{-1}$ was reached in 250~s, \emph{i.e.} a NEA of $3\times10^{-8}$~cm$^{-1}$~Hz$^{-1/2}$. The limit of detection (LOD) in these experiments was quoted as an upper limit because it could not be rigourously assessed due to instabilities in the BPEB source at low concentrations.

Here, we show that CELIF, using a standard UV pulsed dye laser and a modest CRD and LIF setup, can be effective for molecules with fluorescence lifetimes on the order of hundreds of nanoseconds. We present CELIF measurements of the absolute density of SD radicals in a controllable pulsed supersonic jet down to the LOD of $10^5$~cm$^{-3}$. In the 0.002~cm$^3$ probe volume, this corresponds to \emph{ca.}~200 molecules, a quantum-noise-limited $\alpha_\text{min}=7.9\times10^{-11}$~cm$^{-1}$ in 200~s and a NEA of $1.6\times10^{-9}$~cm$^{-1}$~Hz$^{-1/2}$.

\begin{figure}[htbp]
\centering
  \includegraphics[height=5.5cm]{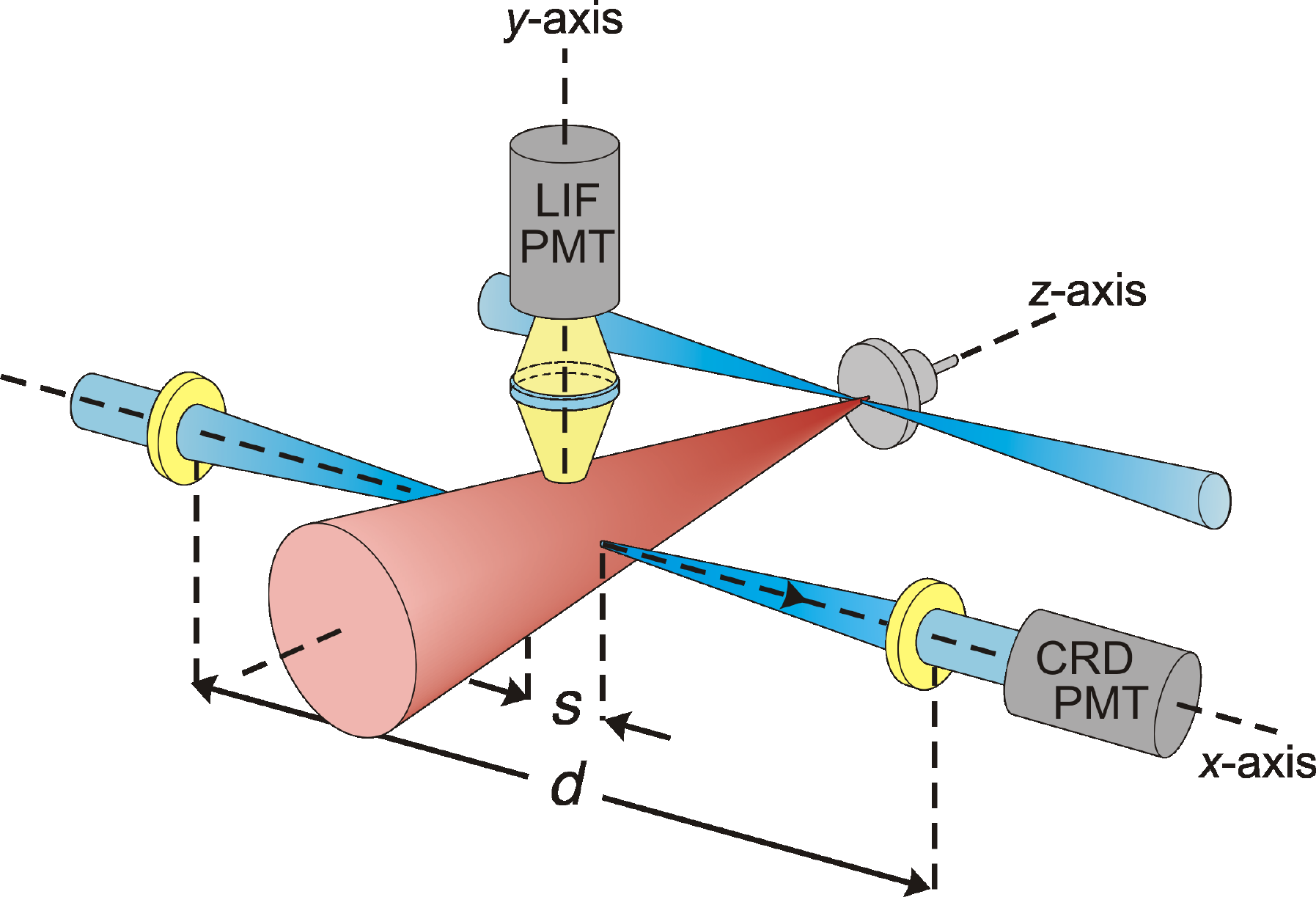}
  \caption{CELIF experimental setup. SD radicals generated by photodissociation of \ce{D2S} in the expansion of the supersonic beam ($z$-axis) are excited by the CRD laser ($x$-axis) and the fluorescence is detected ($y$-axis) simultaneously with the ring-down signal. \label{Fig:setup}}
\end{figure}

Fig.~\ref{Fig:setup} shows a schematic representation of our experiment. A nanosecond pulsed Nd:YAG (Continuum Surelite II-10, 532 nm, repetition rate $f_\text{rep}=10$~Hz, 5~ns pulse length) pumped dye laser (Sirah Cobra Stretch), tuned to the spectrally isolated P$_1(1.5)$ line of the $(0,0)$-band of the $\text{A}^2\Sigma^+\leftarrow\text{X}^2\Pi_{3/2}$ transition at 323.17~nm (200~$\mu$J typical pulse energy), is coupled via beam-shaping, mode-matching and polarisation-rotating optics into a standard ring-down cavity of length $d=98.29\pm0.05$~cm. The cavity mirrors (Layertec) have a measured reflectivity $\geq 99.87\%$, a quoted transmission of 0.01--0.015\% and a radius of curvature of 1~m, which gives a $1/\text{e}^2$ beam waist radius of 0.23~mm. The ring-down transient is measured with a photomultiplier tube (PMT) (Hamamatsu, H7732-10) and digitised with an oscilloscope (LeCroy, WaveRunner 610Zi). The cavity axis, the $x$-axis (see Fig.~\ref{Fig:setup}), intersects orthogonally a supersonic jet of varying densities of SD radicals seeded in Ne (2~bar backing pressure) propagating along the $z$-axis from a pulsed solenoid valve (Parker, General Valve Series 9). The SD radicals are created by photodissociation of \ce{D2S}, mixed at varying partial pressures in the Ne before expansion, using an ArF excimer laser (GAM LASER, EX5, 193~nm, \textit{ca}.\ 4~mJ per pulse). The operating pressure in the chamber was $<10^{-5}$~mbar. The distance between the nozzle and the cavity axis was 12~cm. A three-lens LIF detection optical system is aligned along the $y$-axis and has a field of view that restricts the probe volume to $1.9\times10^{-3}$~cm$^3$ over a length of $s=1.2$~cm. LIF signal photons are counted with a PMT (Hamamatsu, H3695-10) on a second channel of the oscilloscope.

The methodology for CELIF has been described in detail previously.\cite{Sanders2013} Briefly, the LIF signal
\begin{equation}
S^\text{LIF}=I^\text{LIF}\,\alpha\,\Gamma\,g,
\end{equation}
where $I^\text{LIF}$ is the light intensity that has interacted with the molecules within the probe volume, $\alpha=\sigma\rho$ is the absorption coefficient, $\Gamma$ is the fluorescence quantum yield and $\sigma$ is the laser bandwidth corrected absorption cross section. $g$ is an instrument factor discussed below. For highly reflective mirrors ($\approx 100\%$) and low photon loss per cavity pass (\textit{i.e.} $\alpha\,s\ll1$), $I^\text{LIF}\simeq 2I^\text{CRD}/T$, where $I^\text{CRD}$ is the time-integrated CRD intensity and $T$ is the transmission of the cavity exit mirror.\cite{Sanders2013} Therefore, the CELIF signal
\begin{equation}
S^\text{CELIF}=\frac{S^\text{LIF}}{I^\text{CRD}}=\sigma\,\rho\,\Gamma\,\frac{2g}{T}
\label{Eqn:SCELIF}
\end{equation}
is the integrated or photon-counted LIF signal normalised shot-to-shot to the integrated CRD intensity. The factor $g/T$ is difficult to determine because it contains factors that depend on the instrument, \textit{i.e}.\ $g$ is the product of the fraction of fluorescence photons created in the probe volume that hit the LIF PMT, the quantum efficiency of that PMT and a factor quantifying the convolution of the detection system solid angle with the angular distribution of the fluorescence or light scattering process. In order for the CELIF method to deliver absolute absorption coefficients, the instrument dependent factor must be robustly calibrated. Ideally, this would be done by simultaneously measuring LIF and a CRD absorption, as in Sanders \textit{et al.}\cite{Sanders2013} However, the SD density 12~cm from the nozzle, a distance constrained by our vacuum chambers, was too low to measure a CRD absorption. As an alternative, a separate Rayleigh CELIF scattering measurement was used (here done with dry \ce{N2}) leaving all other experimental parameters unchanged. Taking the ratio between the two CELIF measurements gives
\begin{equation}
\frac{S^\text{CELIF}_\text{SD}}{S^\text{CELIF}_\text{\ce{N2}}}=\frac{\sigma_\text{SD}}{\sigma_\text{\ce{N2}}}\frac{\rho_\text{SD}}{\rho_\text{\ce{N2}}}\frac{\Gamma_\text{SD}}{\Gamma_\text{\ce{N2}}}\frac{g}{g'}.
\label{Eqn:Sratio}
\end{equation}
The fluorescence from SD radicals does not necessarily have the same angular distribution as Rayleigh scattering. However, if the angle between the linear polarisation of the probe light and the LIF detection axis is set to the ``magic angle'', $\theta=54.7^\circ$, the difference in the intensities of scattered light and fluorescence light due to the angular distribution vanishes and $g=g'$.\cite{Zare88} For Rayleigh scattering $\Gamma_\text{\ce{N2}}=1$. On the other hand, $\Gamma_\text{SD}\neq1$, because predissociation of the SD radical competes with fluorescence.\cite{Wheeler1997} Therefore, equation~\ref{Eqn:Sratio} can be rearranged to give
\begin{equation}
\alpha_\text{SD}=\sigma_\text{SD}\rho_\text{SD}=\frac{\rho_\text{\ce{N2}}}{S^\text{CELIF}_\text{\ce{N2}}}\frac{\sigma_\text{\ce{N2}}}{\Gamma_\text{SD}}S^\text{CELIF}_\text{SD},
\label{Eqn:alphaSD}
\end{equation}
which no longer contains the instrument dependent factors.

\begin{figure}[t]
\centering
  \includegraphics[height=4.2cm]{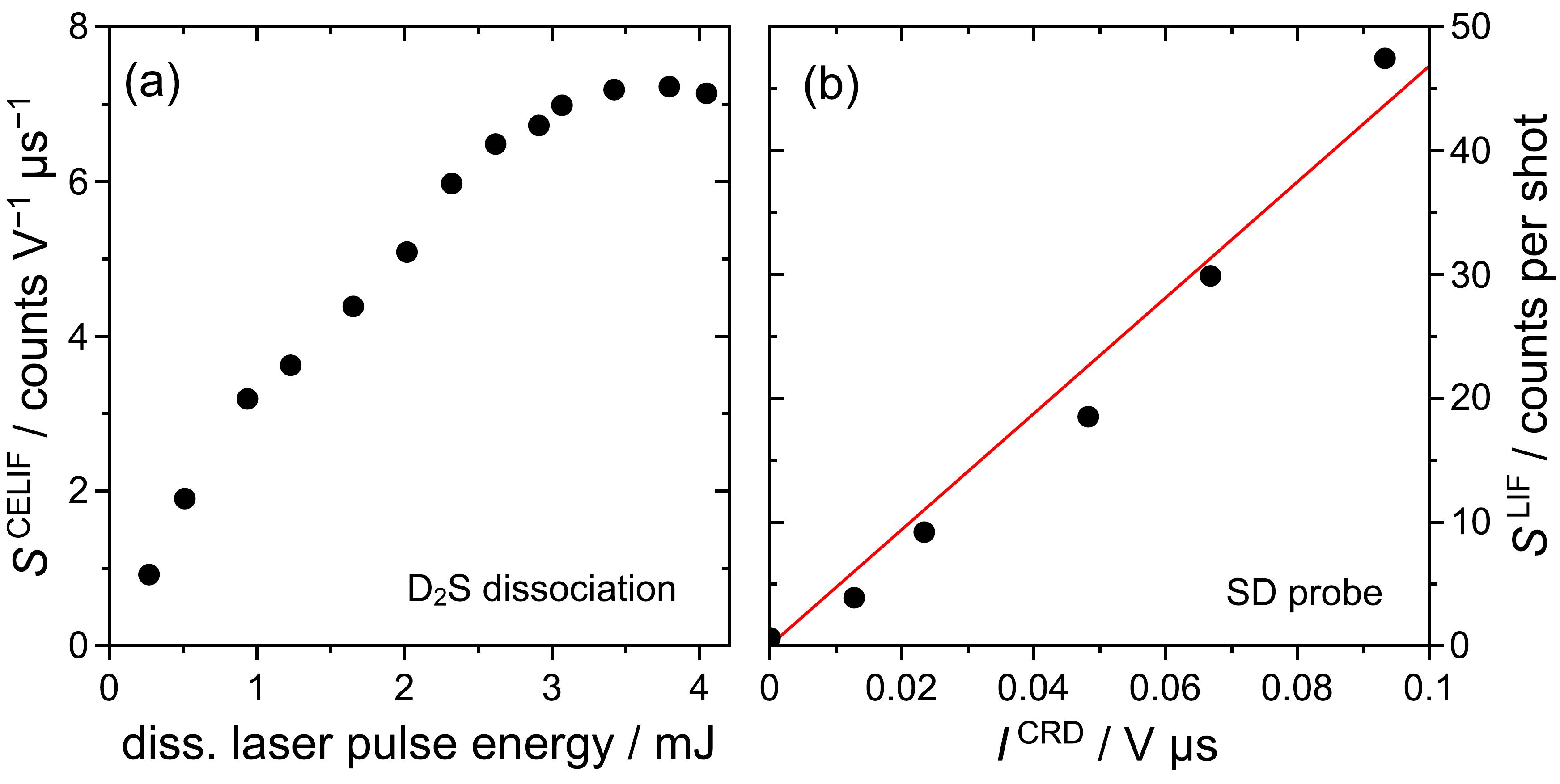}
  \caption{(a) Dependance of the SD CELIF signal on the pulse energy of the \ce{D2S} dissociation laser showing saturation at 3.5 mJ and above. (b) Dependance of the average LIF counts per shot on the probe laser pulse energy as measured by the time-integrated ring-down signal. The statistical error bars are smaller than the symbols. \label{Fig:laserdep}}
\end{figure}

The measurements of $S^\text{CELIF}_\text{SD}$ had to take place in a regime where the photodissociation of the \ce{D2S} molecules was saturated, which ensures that any shot-to-shot instabilities or long-term drifts in the laser pulse energy do not contribute to the noise in $S^\text{CELIF}_\text{SD}$. Fig.~\ref{Fig:laserdep}a shows $S^\text{CELIF}_\text{SD}$ as a function of photodissociation laser pulse energy. The signal varies linearly at low laser pulse energies and begins to enter a saturated regime at around 3.5~mJ per pulse. Unlike the photodissociation process, it is crucial that the fluorescence of SD radicals is \textit{not} saturated for the shot-to-shot normalisation of the CELIF to be valid. Fig.~\ref{Fig:laserdep}b shows that the dependence of $S^\text{CELIF}_\text{SD}$ on the intensity of the probe light in the cavity is linear thus proving that the experiment is not in the LIF saturation regime. The probe laser intensity was reduced sufficiently to ensure that the number of signal counts was not under-represented due to photon coincidences on the detector.

\begin{figure}[t]
\centering
  \includegraphics[height=4.2cm]{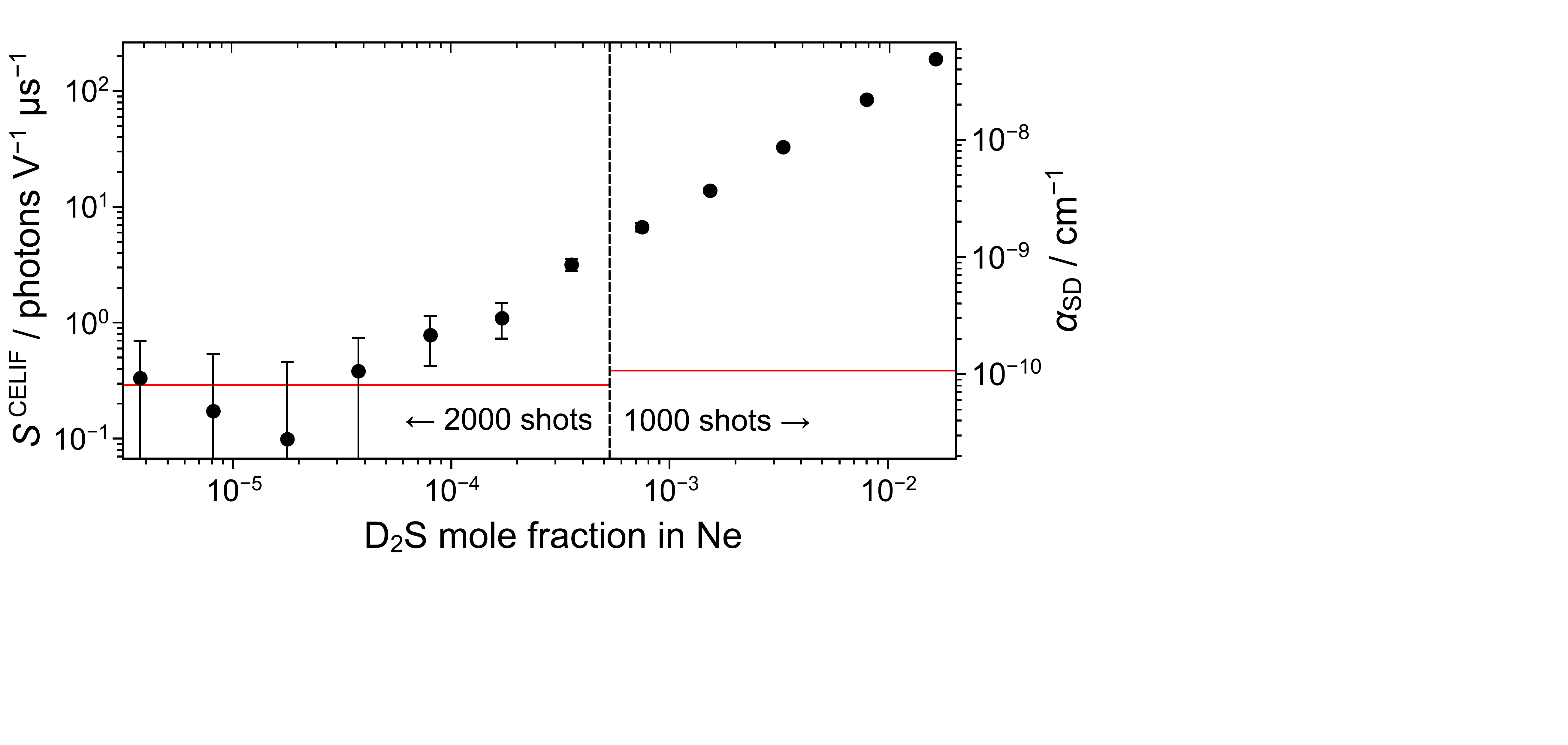}
  \caption{Determination of the limit of detection of the SD CELIF measurement and, therefore, of $\alpha_\text{SD}$ by successive dilution of the \ce{D2S}/Ne gas mixture. \label{Fig:dilution}}
\end{figure}

Fig.~\ref{Fig:dilution} shows a plot of $S^\text{CELIF}_\text{SD}$ versus the mole fraction of \ce{D2S} mixed in the Ne carrier gas before expansion. Starting from the highest value, the mole fraction is lowered by sequentially diluting the previous mixture by pumping away a fraction of the 2~bar total pressure and adding Ne until the total pressure returned to 2~bar. After each dilution, the gas mixture was ``stirred'' for 20~minutes using convection currents generated by heating two loops of pipe of unequal length each entering the bottom and exiting the top of the gas mixing bottle feeding the solenoid valve.

The data manipulation procedure that was used in the analysis is explained in detail in Supplementary Information, Section I. Briefly here, the procedure was as follows. For each dilution the number of photon counts on the LIF PMT $S^\text{LIF}_{\text{tot},i}$ for each $i$th shot was recorded for $n$ laser shots to give the ``total'' signal counts originating from SD molecules and from background sources. Simultaneously, the CRD transient was recorded for each $i$th laser shot and was fitted with an exponential to obtain the amplitude $A^\text{CRD}_{0,i}$, the ring-down time $\tau_i$ and, thus, the integrated CRD intensity using $I^\text{CRD}_{\text{tot},i}=A^\text{CRD}_{0,i}\tau_i$. The total CELIF signal $S^\text{CELIF}_{\text{tot},i}=S^\text{LIF}_{\text{tot,i}}/I^\text{CRD}_{\text{tot},i}$ was calculated for each laser shot and averaged over all laser shots to give the average total CELIF signal per shot $S^\text{CELIF}_\text{tot}$.

Immediately following $n$ laser shots, the photodissociation laser was blocked and the same procedure as above was repeated for another $n$ laser shots to arrive at the average CELIF signal per shot from background sources $S^\text{CELIF}_\text{bg}$. The average CELIF signal per shot originating from SD molecules was thus determined by the background subtraction
\begin{equation}
S^\text{CELIF}_\text{SD}=S^\text{CELIF}_\text{tot}-S^\text{CELIF}_\text{bg}.
\label{Eqn:celifbgsub}
\end{equation}
The CELIF normalisation of the background is possible because 99\% of background photon counts $S^\text{LIF}_\text{bg}$ originated from scattered light from the probe laser (most likely from fluorescence of the UV grade fused silica substrate of the cavity entrance mirror) and was found to be proportional to $I^\text{CRD}_\text{bg}$. The noise in $S^\text{CELIF}_\text{SD}$ (and, therefore, the error bars in Fig.~\ref{Fig:dilution}) is given by
\begin{equation}
\delta S^\text{CELIF}_\text{SD}=\left[\left(\delta S^\text{CELIF}_\text{tot}\right)^2+2\left(\delta S^\text{CELIF}_\text{bg}\right)^2\right]^{1/2},
\end{equation}
where
\begin{equation}
\delta S^\text{CELIF}_\text{tot}=\left[\frac{1}{nS^\text{LIF}_\text{tot}}+\beta(\tau)^2\right]^{1/2}S^\text{CELIF}_\text{tot}
\label{Eqn:celiftoterr}
\end{equation}
and 
\begin{equation}
\delta S^\text{CELIF}_\text{bg}\approx\left(\frac{1}{nS^\text{LIF}_\text{bg}}\right)^{1/2}S^\text{CELIF}_\text{bg}.
\label{Eqn:celifbgerr}
\end{equation}
As explained in Supplementary Information, Section I, the first term in the square root of eqn~(\ref{Eqn:celiftoterr}) is the quantum noise from counting LIF photons over $n$ laser shots and $\beta(\tau)$ is the (approximately) constant relative error in determining the ring-down time from an exponential fit of the CRD transients. $\beta(\tau)$ is neglected in eqn~\ref{Eqn:celifbgerr} because the total noise is dominated by the quantum noise.

The horizontal line in Fig.~\ref{Fig:dilution} represents the limit of detection (LOD) of the CELIF signal where the LIF signal-to-noise ratio is unity. It is calculated using
\begin{equation}
S^\text{CELIF,LOD}_\text{SD}\simeq\left(\frac{2S^\text{LIF}_\text{bg}}{n(I^\text{CRD})^2}\right)^{1/2},
\label{Eqn:celiflod2}
\end{equation}
with $n=1000$ shots and $n=2000$ shots at the higher and lower \ce{D2S} mole fractions, respectively (see Supplementary Information, Section II for derivation).

\begin{figure}[t]
\centering
  \includegraphics[height=5.3cm]{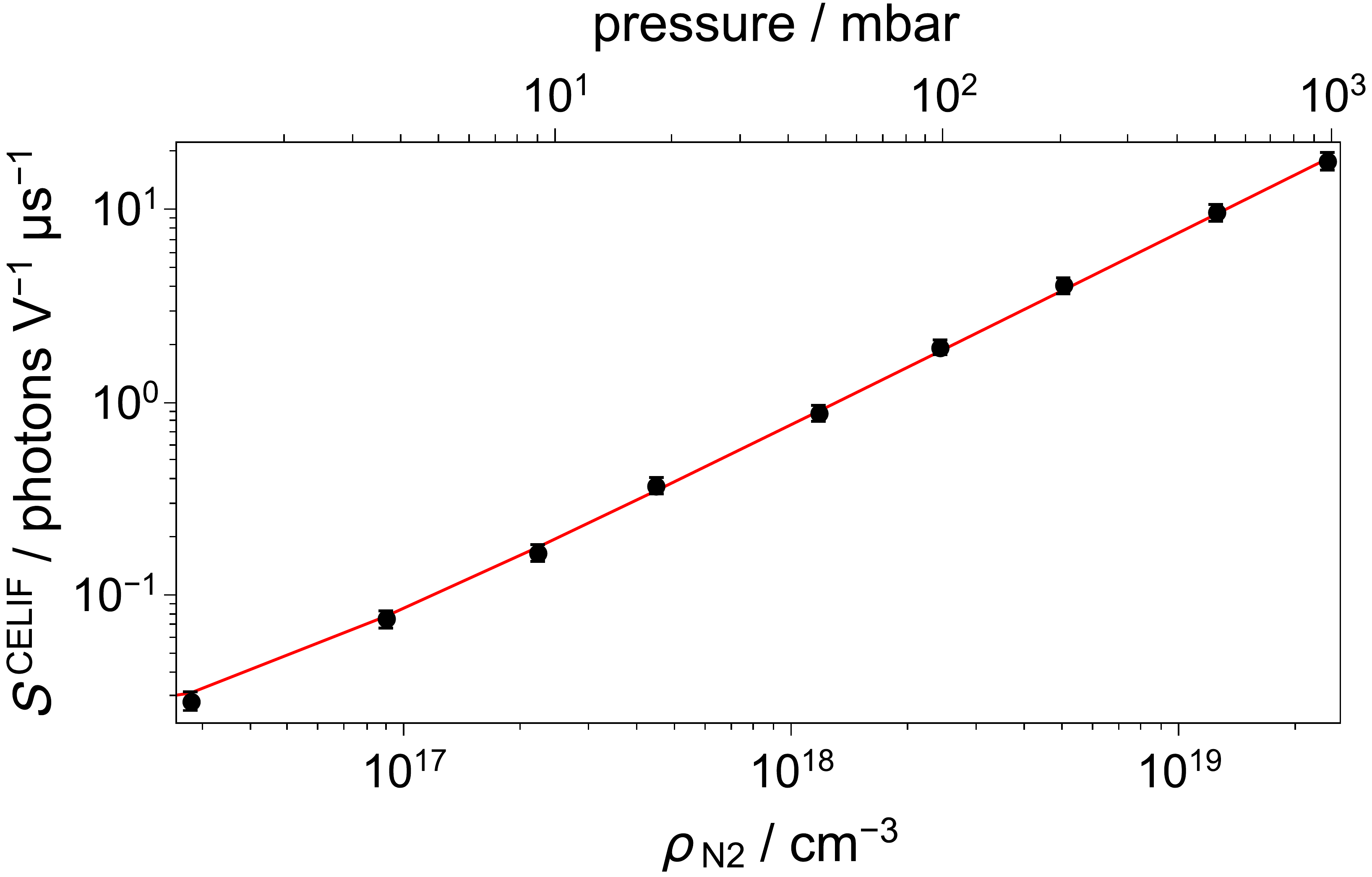}
  \caption{Determination of the ratio $\rho_\text{\ce{N2}}/S^\text{CELIF}_\text{\ce{N2}}$ from measurements of the CELIF signal from \ce{N2} Rayleigh scattering at varying \ce{N2} pressures. The line is a linear $\chi^2$ fit of the data. \label{Fig:N2rayleigh}}
\end{figure}

On Fig.~\ref{Fig:dilution}, the $S^\text{CELIF}_\text{SD}$ axis has been converted into $\alpha_\text{SD}$ using eqn~(\ref{Eqn:alphaSD}). The ratio $\rho_\text{\ce{N2}}/S^\text{CELIF}_\text{\ce{N2}}$ was measured to be $(1.31\pm0.02)\times10^{15}$~counts~per~shot~V~$\mu$s~cm$^{-3}$ from the inverse of the slope of the number density dependence of the \ce{N2} Rayleigh scattering, which is shown in Fig.~\ref{Fig:N2rayleigh}. The slope and the error were determined from a linear $\chi^2$ fit. The same procedure as described above to evaluate $S^\text{CELIF}_\text{SD}$, and associated errors, was also used to evaluate the average \ce{N2} Rayleigh scattering CELIF signal per shot $S^\text{CELIF}_\text{\ce{N2}}$ and its error $\delta S^\text{CELIF}_\text{\ce{N2}}$, with the exception that no background subtraction was required as it did not affect the slope of the graph in Fig.~\ref{Fig:N2rayleigh}.

As per the SD CELIF measurements, the probe laser intensity was reduced sufficiently to ensure good photon counting statistics. The \ce{N2} pressures in the chamber above and below 48~mbar were measured using a calibrated piezo transducer gauge (Pfeiffer Vacuum, APR~265) and a calibrated Pirani gauge (Pfeiffer Vacuum, PBR~260), respectively. Pressures were converted into densities via the van der Waals equation. In principle, the \ce{N2} density could have been measured by CRD. Unfortunately, the alignment of the cavity mirrors changed between the measurement of the reference empty cavity ring-down time and the measurement with \ce{N2} present because of the changing pressure differential. Fortunately, the CELIF technique does not rely on a stable cavity alignment because the normalisation only depends on knowing $I^\text{CRD}$ for a given measurement. To ensure an accurate determination of $I^\text{CRD}$, the cavity was realigned for each \ce{N2} pressure to obtain a good quality exponential decay. 

The minimum detectable absorption coefficient according to eqn~(\ref{Eqn:alphaSD}) is,  $\alpha^\text{LOD}_\text{SD}=(7.9\pm0.6)\times10^{-11}$~cm$^{-1}$, using the values $\sigma_\text{\ce{N2}}(323.17~\text{nm})=(4.1\pm0.2)\times10^{-26}$~cm$^2$,\cite{Ityaksov2008} and $\Gamma_\text{SD}=0.20\pm0.01$.\cite{Wheeler1997,Kawasaki1989} PGOPHER\cite{Western} was used, inputting known spectroscopic constants,\cite{Ramsay52,Senekowitsch85,Klisch96} to obtain a cross section for SD of $(4.5\pm0.2)\times10^{-15}$~cm$^2$~GHz. Using a measured laser bandwidth of $(6.4\pm0.3)$~GHz, the bandwidth corrected cross section is $\sigma_\text{SD}=(7.0\pm0.5)\times10^{-16}$~cm$^2$. The noise at the LOD is dominated by the quantum-noise in the LIF signal, therefore, the quantum-noise-limited NEA\cite{Vanzee1999} for this experiment is given by
\begin{equation}
\text{NEA}=\alpha^\text{LOD}_\text{SD}\left(\frac{2n}{f_\text{rep}}\right)^{1/2}=(1.6\pm0.1)\times10^{-9}~\text{cm}^{-1}~\text{Hz}^{-1/2}.
\label{Eqn:nea}
\end{equation}
The density of SD radicals at our LOD is $\rho^\text{LOD}_\text{SD}=1.1\times10^5$~molecules~cm$^{-3}$ and has an error of only 10\%. Given the very small probe volume, this density corresponds to  an average of just $215\pm 21$ molecules in the probe volume at any given time. 

At the heart of CELIF is a fluorescence measurement. Therefore, it should be noted that the technique is limited to molecules that are fundamentally suitable for LIF. In addition, molecules must fluoresce on a timescale less than the time it takes for a molecule to leave the field of view of the LIF detection optics.\footnote{For the SD molecules in these experiments, the fluorescence lifetime is \emph{ca.}~200~ns and the time the molecules take to leave the field of view of the LIF detection optics from the point of excitation is around 7~$\mu$s.} Also, the molecule must absorb in a wavelength region $>200$~nm where CRD mirrors can be fabricated. However, many important molecules of interest in many fields fall into this category (\textit{e.g.} NO). 

The work of Sanders \textit{et al.}\ did not prove that CELIF would be as effective as it was for BPEB at measuring the absolute absorption coefficients of molecules with relatively long fluorescence lifetimes, such as SD.\cite{Sanders2013} Here we have shown that CELIF is effective. These measurements are potentially of use to those wishing to determine absolute collision cross sections in molecular beam scattering experiments where knowledge of densities is required.\cite{Kirste2012}

Like cavity enhanced absorption techniques, CELIF can measure absorption coefficients, but via a fluorescence measurement. Therefore, it is useful to place these UV measurements on SD in the context of other absolute measurements where absorption coefficients are obtained. The NEA in these CELIF experiments is three orders of magnitude better than the UV measurements of Jongma \textit{et al.} Compared to absolute measurements in the visible by Osthoff \textit{et al.}, the CELIF NEA here is one order of magnitude worse.

Given the modest setup used in these experiments, there is much room to improve the LOD. Examining eqn~(\ref{Eqn:celiflod2}) reveals that maximising $n$ and $I^\text{CRD}$ and/or minimising $S^\text{LIF}_\text{bg}$ will lower the LOD of the experiment. Even though the cavity is an effective discriminator of scattered light compared to a standard LIF setup, 99\% of $S^\text{LIF}_\text{bg}$ originated from scattered light from the probe laser, as explained earlier. Thin mirrors with a low absorbance allied with light baffles and low reflectivity surfaces inside the cavity chamber, as is done routinely in standard LIF measurements, could effectively eliminate the scattered light leaving only dark counts on the PMT as the source of background signal. A photon-counting PMT (\textit{e.g}.\ Hamamatsu H7360-01, 2~pA dark current) could lower the dark counts by a further factor of $10^3$. The result would thus be to lower the NEA to a value on the order of $10^{-12}~\text{cm}^{-1}~\text{Hz}^{-1/2}$ and the LOD density to \textit{ca.}~400~cm$^{-3}$ (on average less than 1 molecule in the probe volume at any given time). Using a narrow bandwidth laser of, say, 120~MHz\cite{Kirste2012} would not change the NEA, but would increase the bandwidth corrected SD absorption cross-section by a factor of 50 and commensurately lower the LOD density to $<10$~cm$^{-3}$.

This work was supported by the EPSRC [grant no. EP/I012044/1]. NHN thanks the Royal Society for a University Research Fellowship. The authors would like to thank C. M. Western, G. Meijer, S. Y. T. van de Meerakker, J. R. R. Verlet, T. Momose and S. Willitsch for useful discussions.

\footnotesize{
\bibliography{rsc} 
\bibliographystyle{rsc} }

\section*{Supplementary Information}

\subsection*{Data and error analysis}

The number of photon counts on the LIF PMT $S^\text{LIF}_{\text{tot},i}$ for each $i$th shot was recorded for $n$ laser shots to give the ``total'' signal counts originating from SD molecules and from background sources. Simultaneously, the cavity ring-down (CRD) transient was recorded for each $i$th laser shot. Each CRD transient was fitted with an exponential to obtain the amplitude $A^\text{CRD}_{0,i}$ and the ring-down time $\tau_i$ and the integrated CRD intensity calculated
\begin{equation}
I^\text{CRD}_{\text{tot},i}=A^\text{CRD}_{0,i}\tau_i.
\end{equation}
The total CELIF signal for each laser shot
\begin{equation}
S^\text{CELIF}_{\text{tot},i}=\frac{S^\text{LIF}_{\text{tot,i}}}{I^\text{CRD}_{\text{tot},i}}
\end{equation}
was then calculated and averaged over all laser shots to give the average total CELIF signal per shot $S^\text{CELIF}_\text{tot}$. This procedure was found to give the same result as using
\begin{equation}
S^\text{CELIF}_\text{tot}=\frac{S^\text{LIF}_\text{tot}}{I^\text{CRD}_\text{tot}},
\end{equation}
where the subscript $i$ has been dropped because $S^\text{LIF}_\text{tot}$ is the average number of photon counts per laser shot and $I^\text{CRD}_\text{tot}$ is the average integrated CRD intensity per shot.\\

Immediately following $n$ laser shots, the dissociation laser was blocked and the same procedure above was repeated for another $n$ laser shots to arrive at the average CELIF signal per shot from background sources $S^\text{CELIF}_\text{bg}$. The average CELIF signal per shot originating from SD molecules was thus determined by the background subtraction
\begin{equation}
S^\text{CELIF}_\text{SD}=S^\text{CELIF}_\text{tot}-S^\text{CELIF}_\text{bg}
\label{Eqn:bgsub}
\end{equation}
and is plotted in Fig.~3. The CELIF normalisation of the background is justified when one considers the sources of the background signal. Even though the cavity is an effective discriminator of scattered light compared to a standard LIF setup, 99\% of $S^\text{LIF}_\text{bg}$ originated from scattered light from the probe laser, most likely from UV fluorescence of the UV grade fused silica substrate of the cavity entrance mirror. $S^\text{LIF}_\text{bg}$ was found to be proportional to $I^\text{CRD}_\text{bg}$.
\\

The error bars in Fig.~3 for $S^\text{CELIF}_\text{SD}$ were determined as follows. The relative fitting error on $A^\text{CRD}_{0,i}$ and $\tau_i$, was
\begin{equation}
\beta(A_{0,i})=\frac{\delta A^\text{CRD}_{0,i}}{A^\text{CRD}_{0,i}}=0.1\%~~~~~\text{and}~~~~~\beta(\tau_i)=\frac{\delta\tau_i}{\tau_i}=1\%,
\end{equation}
respectively. The relative fitting errors, $\beta(A_{0,i})$ and $\beta(\tau_i)$, were found on analysis to vary insignificantly over all laser shots because the cavity was set up such that the ring-down transients were good quality single exponential decays with electronic noise that did not vary significantly shot-to-shot. Thus, $\beta(A_{0,i})$ and $\beta(\tau_i)$ are treated as constants, $\beta(A_0)$ and $\beta(\tau)$, where $\beta(A_0)\ll \beta(\tau)$. The resulting relative error $\beta(I)$ on the \emph{determination} of each $I^\text{CRD}_{\text{tot},i}$ only depends on the fitting errors and is, therefore
\begin{equation}
\beta(I)=\sqrt{\beta(A)^2+\beta(\tau)^2}\approx \beta(\tau),
\end{equation}
$\beta(I)$ is the only relevant error with respect to $I^\text{CRD}_{\text{tot},i}$ because an important point of CELIF is to remove noise caused by shot-to-shot fluctuations in laser intensity. The quantum noise from counting LIF photons over $n$ laser shots was
\begin{equation}
\delta S^\text{LIF}_\text{tot}=\left(\frac{S^\text{LIF}_\text{tot}}{n}\right)^{1/2},
\end{equation}
and comes from Poisson statistics. Therefore, the combined noise in $S^\text{CELIF}_\text{tot}$ is
\begin{equation}
\delta S^\text{CELIF}_\text{tot}=\left[\frac{1}{nS^\text{LIF}_\text{tot}}+\beta(\tau)^2\right]^{1/2}S^\text{CELIF}_\text{tot}.
\label{Eqn:sceliftoterror}
\end{equation}
An equivalent to eqn~(\ref{Eqn:sceliftoterror}) for the background CELIF signal can be derived in the same way except the error is dominated by the quantum noise because $S^\text{LIF}_\text{bg}$ is very small, \emph{i.e.}
\begin{equation}
\delta S^\text{CELIF}_\text{bg}\approx\left(\frac{1}{nS^\text{LIF}_\text{bg}}\right)^{1/2}S^\text{CELIF}_\text{bg}.
\label{Eqn:scelifbgerror}
\end{equation}
The noise in $S^\text{CELIF}_\text{SD}$ from eqn (\ref{Eqn:bgsub}) is
\begin{equation}
\delta S^\text{CELIF}_\text{SD}=\left[\left(\delta S^\text{CELIF}_\text{tot}\right)^2+2\left(\delta S^\text{CELIF}_\text{bg}\right)^2\right]^{1/2},
\end{equation}
where the factor 2 comes from the fact that $S^\text{CELIF}_\text{tot}$ contains $S^\text{CELIF}_\text{bg}$.

\subsection*{Limit of detection derivation}

To derive an expression for the limit of detection (LOD) of $S^\text{CELIF}_\text{SD}$, which is the horizontal line in Fig.~3, the approximation that $I^\text{CRD}_\text{tot}=I^\text{CRD}_\text{bg}=I^\text{CRD}$ is made such that
\begin{equation}
S^\text{LIF}_\text{SD}=S^\text{LIF}_\text{tot}-S^\text{LIF}_\text{bg},
\end{equation}
so the noise in $S^\text{LIF}_\text{SD}$ is
\begin{equation}
\left(\delta S^\text{LIF}_\text{SD}\right)^2=\left(\delta S^\text{LIF}_\text{tot}\right)^2+2\left(\delta S^\text{LIF}_\text{bg}\right)^2=\frac{1}{n}\left(S^\text{LIF}_\text{SD}+2S^\text{LIF}_\text{bg}\right).
\end{equation}
At the LOD, $S^\text{LIF}_\text{SD}=S^\text{LIF,LOD}_\text{SD}=\delta S^\text{LIF}_\text{SD}$, therefore
\begin{equation}
n\left(S^\text{LIF,LOD}_\text{SD}\right)^2=S^\text{LIF,LOD}_\text{SD}+2S^\text{LIF}_\text{bg}.
\end{equation}
Solving for $S^\text{LIF,LOD}_\text{SD}$ gives
\begin{equation}
S^\text{LIF,LOD}_\text{SD}=\frac{1}{2n}+\left(\frac{1}{4n^2}+\frac{2S^\text{LIF}_\text{bg}}{n}\right)^{1/2},
\end{equation}
which, in the limit of large $n$, gives
\begin{equation}
S^\text{LIF,LOD}_\text{SD}\approx\left(\frac{2S^\text{LIF}_\text{bg}}{n}\right)^{1/2}.
\end{equation}
The CELIF signal at the LOD is therefore
\begin{equation}
S^\text{CELIF,LOD}_\text{SD}\approx\left(\frac{2S^\text{LIF}_\text{bg}}{n\left(I^\text{CRD}\right)^2}\right)^{1/2},
\end{equation}
which, when compared to eqn~(\ref{Eqn:scelifbgerror}), reveals that
\begin{equation}
S^\text{CELIF,LOD}_\text{SD}\approx\sqrt{2}\delta S^\text{CELIF}_\text{bg}.
\end{equation}

\end{document}